\documentclass[prb,preprint,showpacs]{revtex4}
\usepackage{graphics}
\begin{document}
\title{Enhancement of low field 
magnetoresistance at room temperature in  
La$_{0.67}$Sr$_{0.33}$MnO$_{3}$/Al$_{2}$O$_{3}$ nanocomposite}
\author{Soumik Mukhopadhyay}
\author{I. Das}
\affiliation{Saha Institute of Nuclear Physics, 1/AF, Bidhannagar, Kolkata 
700064, India}
\begin{abstract}
Magnetotransport properties in a nanocrystalline 
La$_{0.67}$Sr$_{0.33}$MnO$_{3}$/micron sized Al$_{2}$O$_{3}$ granular composite
with different concentrations of Al$_{2}$O$_{3}$ have been studied.
The resistivity curves in absence of magnetic field and the various transport
mechanisms which might account for the upturn in resistivity 
at low temperature, has been discussed.
Enhancement of low field magnetoresistance at room temperature with the 
introduction of Al$_{2}$O$_{3}$ has been observed. 
\end{abstract}
\maketitle

\noindent
Study of granular metal-insulator composites got a real boost in the 
early seventies when Sheng
and coworkers~\cite{sheng,sheng1} published a series of papers where they
explained the transport properties of granular nonmagnetic metal-insulator
composites. Helman et al.~\cite{abeles} extended the study to granular 
ferromagnetic
metals in $1976$. In $1998$, Mitani et al~\cite{mitani} explained the enhanced
magnetoresistance in granular ferromagnetic films considering the grain size
distribution and the resulting higher order spin dependent tunneling.
Milner et al.~\cite{composite} pointed out the remarkably similar 
magnetotransport properties of ferromagnet-metal and ferromagnet-insulator 
composites. Besides the technological importance, granular ferromagnets
are very attractive from fundamental point of view in the sense that one
can study the interplay between different interesting phenomena like coulomb 
charging,
spin dependent tunneling etc. Recently there is a renewed interest in the
field of granular ferromagnets with the increasing popularity of novel magnetic
materials like CrO$_{2}$ and perovskite Manganites with extremely high
spin polarization and the possibility of 
enhanced spin dependent transport. So far as the perovskites manganites are
concerned, La$_{0.67}$Sr$_{0.33}$MnO$_{3}$ (LSMO) has the highest 
reported T$_{C}$
compared to it's other counterparts. However the spin polarization decreases
drastically with increasing temperature and hence the achievement of low field 
magnetoresistance at room temperature is a challenging task. 
We have tried to combine the effect
of high degree of spin polarization of LSMO and the characteristics of a 
heterogeneous granular structure to achieve enhancement in magnetoresistance.
Al$_{2}$O$_{3}$ (ALO) is an inert insulator and a popular choice as the 
insulator
in magnetic tunnel junction devices. In $2001$, there was a report on 
enhancement of magnetoresistance in nanocrystalline
La$_{0.67}$Ca$_{0.33}$MnO$_{3}$/AL$_{2}$O$_{3}$ composite~\cite{hueso}. 
However the enhancement was reported to be at low temperature. 
This articles reports the observation of enhanced low field 
magnetoresistance in a nanocrystalline LSMO/ microcrystalline ALO composite
at room temperature.
%\section{Experiment}

\indent
The LSMO nanoparticles were prepared in the powder form using the sol-gel
method with citric acid as the gellifying agent and subsequently heated 
at $1000^{0}$C for $4$ hrs. The crystallinity of the powder was analyzed 
by x-ray diffraction (XRD) (Fig:~\ref{fig:xray}) and the grain size with it's 
distribution was measured using transmission electron 
microscopy (TEM). The average size of the LSMO grains is 
about $50$ nm. The LSMO powder obtained from solgel method followed by
heat treatment was mixed thoroughly with commercially obtained crystalline 
ALO powder at different weight concentrations $x$ 
(where $x = 0, 0.1, 0.2, 0.3, 0.4, 0.5$) of ALO and pelletized. The
pellets were given a  heat treatment at $1000^{0}$C (the 
same temperature at which the sol-gel derived LSMO nanoparticles were
annealed) for $2$ hr. The microstructures of the composites were
observed using a Scanning electron microscope (SEM) (Fig:~\ref{fig:sem}). 
Energy dispersive
x-ray analysis was carried out to analyse the chemical composition 
of the samples. Analysis of x-ray diffraction patterns (Fig:~\ref{fig:xray}) 
for the 
LSMO powder, the ALO powder and the pelletized composite shows that all the
characteristic reflection peaks of LSMO and ALO are present in the composite.
Moreover, the positions of the respective chracteristic x-ray diffraction 
lines do not shift indicating that there is no chemical reaction at the 
interfaces between LSMO and ALO. There is no appreciable change 
in the width of the 
diffraction peaks as well suggesting that the increase in sizes of the LSMO 
particles (after the heat treatment of the pellets) is negligible.  
The magnetotransport properties were studied using four probe method in
the temperature range $2-300$ K.
The magnetic field was applied in the plane of the applied electric field.
%\begin{figure}
%\resizebox {8cm}{9cm}
%{\includegraphics{aloxray.eps}}
%\caption{X-ray diffraction data for LSMO/ALO composite in pellet form with
%50\% wt. concentration along with the parent LSMO and ALO powder. All the
%characteristic peaks of LSMO and ALO are present in the composite system
%and appearence of additional peaks is not observed.}
%\end{figure}
%\begin{figure}
%\resizebox{4.5cm}{4.3cm}
%{\includegraphics{alox0b.EPS}}
%\caption{SEM picture for the sample with ALO concentration $x=0$
%over an area $1 \times 1 \mu m^{2}$}
%\end{figure}
%\begin{figure}
%\resizebox {4.2cm}{4.2cm}
%{\includegraphics{aloresfit.EPS}}
%\resizebox {4.2cm}{4.2cm}
%{\includegraphics{aloresperc.EPS}}
%\caption{On the left: Resistivity curves for (LSMO)$_{1-x}$(ALO)$_{x}$ 
%in absence of magnetic
%field along with the corresponding theoritical fits. The open symbols are
%the experimental data and the continuous lines are the theoritical fits.
%On the right: The resistivity as a function of x at $3$ K and $100$ K}
%\end{figure}

\indent
The resistivity vs. temperature curves (Fig:~\ref{fig:res}A) 
in absence of magnetic field for the samples
(LSMO)$_{1-x}$(ALO)$_{x}$ (where $x = 0, 0.1, 0.2, 0.3, 0.4, 0.5$) were
studied. The metal-insulator transition temperature in the 
resistivity curves shifts systematically towards lower temperature
as ALO concentration increases. This is understandable since
increasing ALO concentration hinders metallic transport and hence the
shift towards lower temperature. There is an upturn in resistivity at low
temperature which gets more pronounced with increasing $x$. 
The resistivity curves can be explained on the basis of a parallel channel
conduction model~\cite{parallel} in which one channel is taken as a metallic 
path and the other as a thermally activated conduction path. 
However, it alone cannot explain the low temperature upturn in
resistivity with decreasing temperature. The average size of the
LSMO nanoparticles being $50$ nm, one cannot rule out 
coulomb charging effect playing an important role at low temperature.
The parallel channel
conduction model can explain the resistivity curves over the whole
range of temperature if we include a coulomb blockade term in the
metallic channel. The coulomb blockade term takes care of the upturn
in resistivity at low temperature. The coulomb charging energy $E_{C}$ 
depends not only on the grain size but also on the microstructure around the 
metallic grains, which might be different for 
different concentrations of ALO. The temperature ($T$) dependence of 
the resistivity due to coulomb charging effect is given by 
$\rho_{C}\propto\exp{b/T^{1/2}}$, where $b$ is a positive constant.
However, even if we include a thermally activated conduction term
like $\rho_{a}\propto\exp(\Delta/KT)$ in the metallic channel, 
the fitting is equally good. Such a term could arise due to a 
small activation barrier observable at low temperature~\cite{parallel}
or due to classical Poole-frenkel~\cite{frenkel} type emission where
the localized charge carriers are thermally activated over the potential
barrier. The electrical resistivity arising out of Poole-Frenkel emission
is given by, $\rho_{PF}=\rho_{0}\exp\left(\beta_{PF}E^{1/2}/KT\right)$
where $\beta_{PF}$ is a constant and $E$ is the applied electric field. 
The electric field dependence of conductivity establishes that indeed
Poole-Frenkel emission is at play. A detailed discussion on this matter
is, however, out of scope for this article. 
The expression for resistivity according to the parallel channel
conduction model is given by,
$1/{\rho}=A/{\rho_{M}}+B/{\rho_{S}}$
where $\rho_{M}$ is the metallic resistivity part
and $\rho_{S}$ is the thermally activated conduction channel represented
by a semiconductor like temperature dependence of resistivity.
However in our case the modified form of the parallel channel conduction
model is given by
\[\frac{1}{\rho}=\frac{A}{\rho_{M}+A_{1}\rho_{PF}}+\frac{B}{\rho_{S}}\]

\indent
The metallic percolation thershold can be identified just
by analysing the resistivity data i.e. to identify the transition
point at which the temperature coefficient of resistivity changes sign
with variation in $x$. However in our case although the metal-insulator 
transition temperature shifts towards lower temperature with increasing $x$, 
the metallic conduction is not
completely suppressed as metal-insulator transition is observed even 
for $x=0.5$. This suggests that percolation threshold for the metallic
grains is still not achieved at $x=0.5$. The percolation threshold for
the insulating grains could possibly be found out
by identifying the point at which the resistivity
at a certain temperature undergoes a more rapid increase with increase 
in $x$. We have plotted the reduced resistivity at temperatures $3$ K 
and $100$ K as a function of $x$ (Fig:~\ref{fig:res}B) 
and observed an almost exponential increase in
resistivity with increase of $x$ at both the temperatures. One can make
a rough idea from the above plot that the percolation threshold for 
the alumina grains is possibly around $x=0.1$ above which the resistivity 
rises more rapidly.
%\begin{figure}
%\resizebox {6.5cm}{5.5cm}
%{\includegraphics{alomrh300K.EPS}}
%\resizebox {6.5cm}{5.5cm}
%{\includegraphics{alomrt5kOe.EPS}}
%\caption{Above: The magnetoresistance as a function of magnetic field for 
%different values of $x=0, 0.1, 0.2, 0.4$ at room temperature. 
%Enhancement of low field room temperature 
%magnetoresistance is observed with introduction of ALO.
%Below: The temperature dependence of magnetoresistance for
%$x=0, 0.1, 0.2, 0.5$ at $H = 5$ kOe. Enhancement of magnetoresistance is
%observed with introduction of ALO near room temperature but at low
%temperature, the enhancement is not observed. The curves corresponding
%to other samples are not shown for clarity.}
%\end{figure}

\indent
The magnetoresistive properties are very interesting. Usually the
enhancement in magnetoresistance is observed around the percolation
threshold for the insulating grains. Formation of percolation pathway 
for the alumina grains leads to enhanced spin dependent transport across
the ferromagnetic metallic grains. 
There are a few reports concerning magnetotransport properties in 
manganite/alumina composites. Hueso et. al.~\cite{hueso} observed 
enhancement in magnetoresistance at $77$ K
for La$_{0.67}$Ca$_{0.33}$MnO$_{3}$/Al$_{2}$O$_{3}$ composite with 
the introduction of alumina. There is another study on LSMO/ALO composite thin 
film~\cite{thin} which also reports the enhancement in magnetoresistance 
only at low temperature.
In the above case the insulator ALO was found out to be 
amorphous in nature. But nothing whatsoever is known about the 
magnetoresistive properties at still higher temperature.
Till date, no enhancement in low field magnetoresistance at room
temperature with the introduction of alumina has been observed in LSMO/ALO
nanocomposite. We have observed enhancement 
in low field magnetoresistance $[\{\rho(H)-\rho(0)\}/\rho(0) (\%)]$ in
manganite/alumina composite at room temperature (Fig:~\ref{fig:mr}A).
The MR peaks at $x=0.1$ which is consistent with our idea about the percolation
threshold for the alumina grains and the enhancement is nearly $100\%$ 
(two times) for $x=0.1$ compared to $x=0$ at $200$ Oe. The percolation 
threshold depends not only on the concentration of the insulator but also
on the packing fraction of the pellet. The low value of the percolation
threshold can be understood taking into consideration the packing fraction
and the tendancy of the ferromagnetic metallic grains to form chain-like
structure due to dipole-dipole interactions. The temperature dependence of
magnetoresistance at $5$ kOe (Fig:~\ref{fig:mr}B) reveals that the 
enhancement in MR with
the introduction of alumina is achieved near room temperature
whereas at low temperature the low field MR decreases with the increse
of $x$. The behaviour at low temperature is unusual and at the moment remains
unexplained. For $x=0.1$ the enhancement is sustained down to $200$ K
whereas for $x=0.2$ the enhancement is sustainable within the temperature
range $250-300$ K. 

\indent
To summarize, magnetotransport properties have been studied
in nanocrystalline LSMO/microcrystalline ALO composite.
Enhancement in low field magnetoresistance has been
achieved at room temperature with the introduction of alumina.

\newpage
\begin{figure}
%\resizebox {8cm}{9cm}
%{\includegraphics{aloxray.eps}}
\caption{X-ray diffraction data for LSMO/ALO composite in pellet form with
50\% wt. concentration along with the parent LSMO and ALO powder. All the
characteristic peaks of LSMO and ALO are present in the composite system
and appearence of additional peaks is not observed.}\label{fig:xray}
%\end{figure}
%\begin{figure}
%\resizebox{4.5cm}{4.3cm}
%{\includegraphics{alox0b.EPS}}
\caption{SEM picture for the sample with ALO concentration $x=0$
over an area $1 \times 1 \mu m^{2}$}\label{fig:sem}
%\end{figure}
%\begin{figure}
%\resizebox {4.2cm}{4.2cm}
%{\includegraphics{aloresfit.EPS}}
%\resizebox {4.2cm}{4.2cm}
%{\includegraphics{aloresperc.EPS}}
\caption{{\textbf A}: Resistivity curves for (LSMO)$_{1-x}$(ALO)$_{x}$
in absence of magnetic
field along with the corresponding theoritical fits. The open symbols are
the experimental data and the continuous lines are the theoritical fits.
{\textbf B}: The resistivity as a function of x at $3$ K and $100$ K}
\label{fig:res}
%\end{figure}
%\begin{figure}
%\resizebox {6.5cm}{5.5cm}
%{\includegraphics{alomrh300K.EPS}}
%\resizebox {6.5cm}{5.5cm}
%{\includegraphics{alomrt5kOe.EPS}}
\caption{{\textbf A}: The magnetoresistance as a function of magnetic field for
different values of $x=0, 0.1, 0.2, 0.4$ at room temperature.
Enhancement of low field room temperature
magnetoresistance is observed with introduction of ALO.
{\textbf B}: The temperature dependence of magnetoresistance for
$x=0, 0.1, 0.2, 0.5$ at $H = 5$ kOe. Enhancement of magnetoresistance is
observed with introduction of ALO near room temperature but at low
temperature, the enhancement is not observed. The curves corresponding
to other samples are not shown for clarity.}
\label{fig:mr}
\end{figure}
$~~~~~~~~~~~$
\vskip 4.5in
\newpage

\begin{figure}
\resizebox {8cm}{9cm}
{\includegraphics{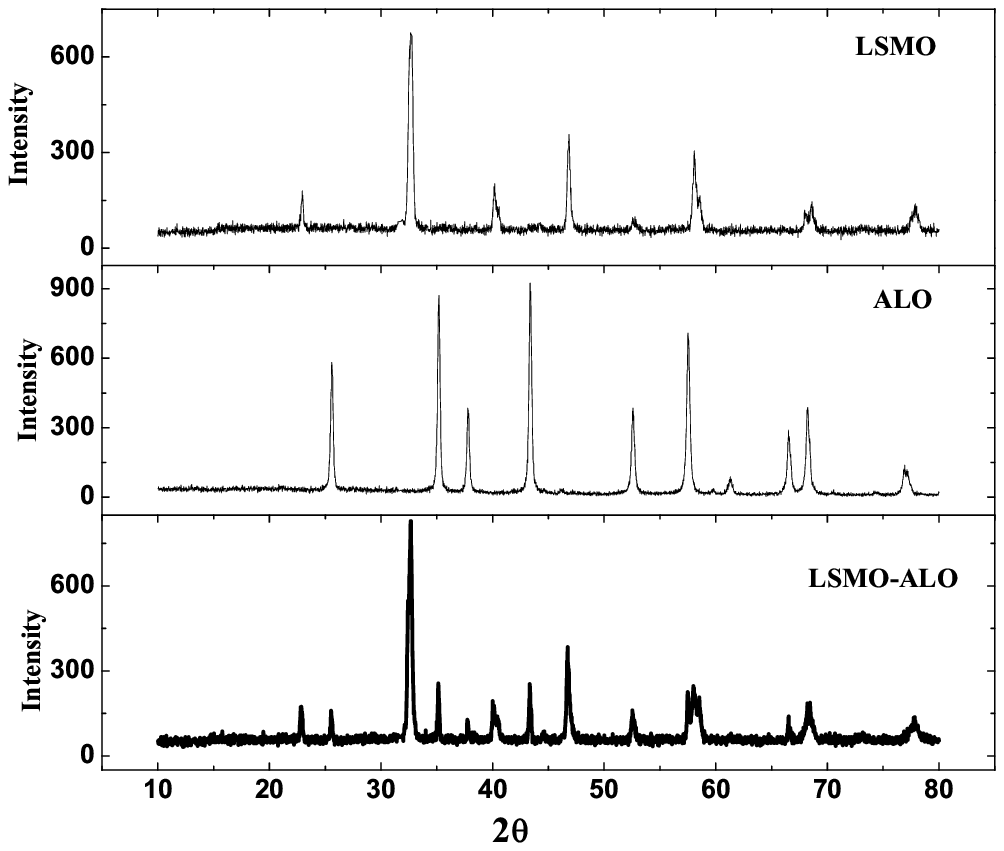}}
%\caption{
%X-ray diffraction data for LSMO/ALO composite in pellet form with
%50\% wt. concentration along with the parent LSMO and ALO powder. All the
%characteristic peaks of LSMO and ALO are present in the composite system
%and appearence of additional peaks is not observed.
%}
%\label{fig:xray}
\end{figure}
\centering{Fig:1}
$~~~~~~~~$
\vskip 4.5in
\newpage
\begin{figure}
\resizebox{4.5cm}{4.3cm}
{\includegraphics{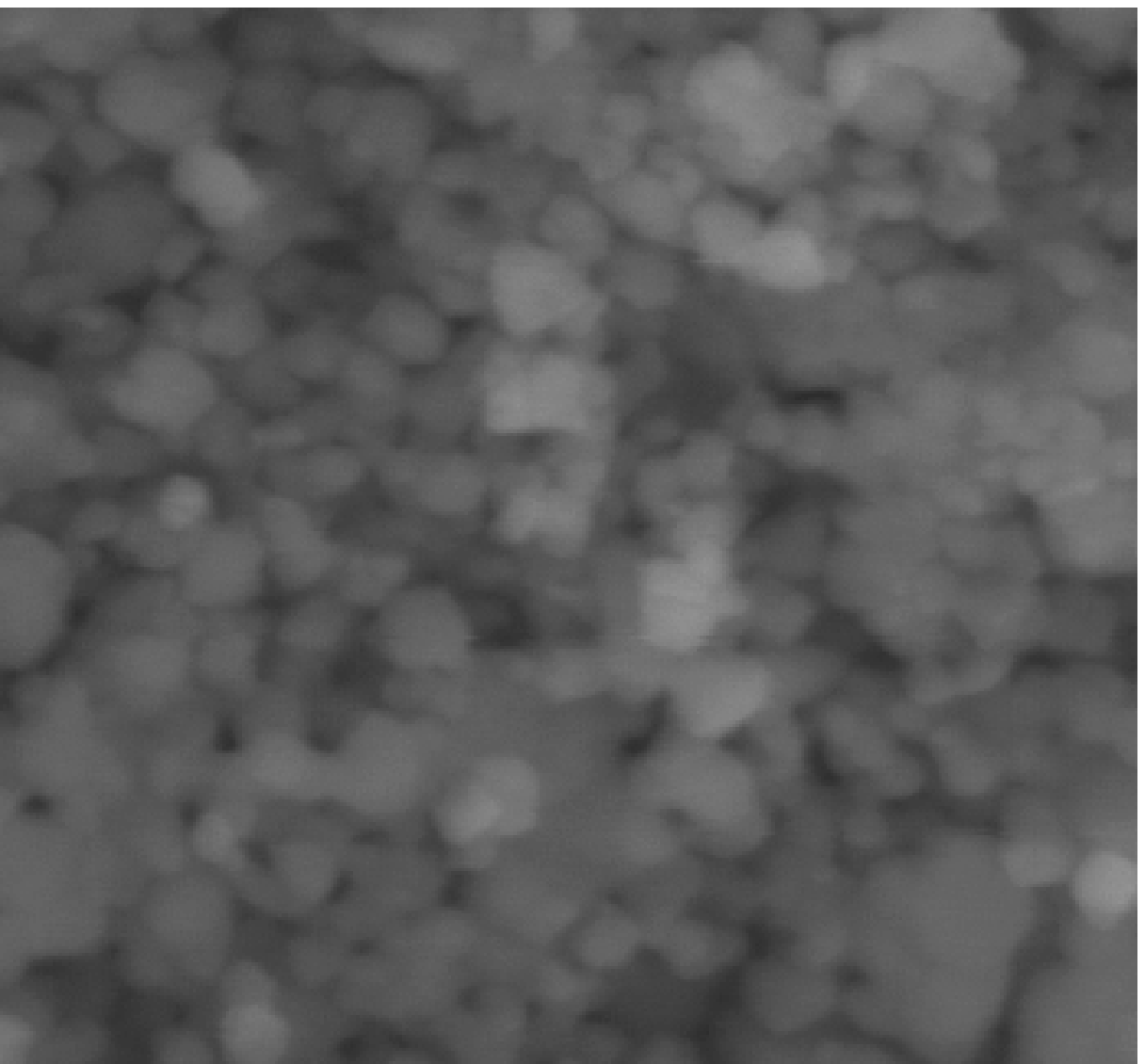}}
%\caption{
%SEM picture for the sample with ALO concentration $x=0$
%over an area $1 \times 1 \mu m^{2}$
%}
%\label{fig:sem}
\end{figure}
\centering{Fig:2}
$~~~~~~~~~~~$
\vskip 4.5in
\newpage
\begin{figure}
\resizebox {4.2cm}{4.2cm}
{\includegraphics{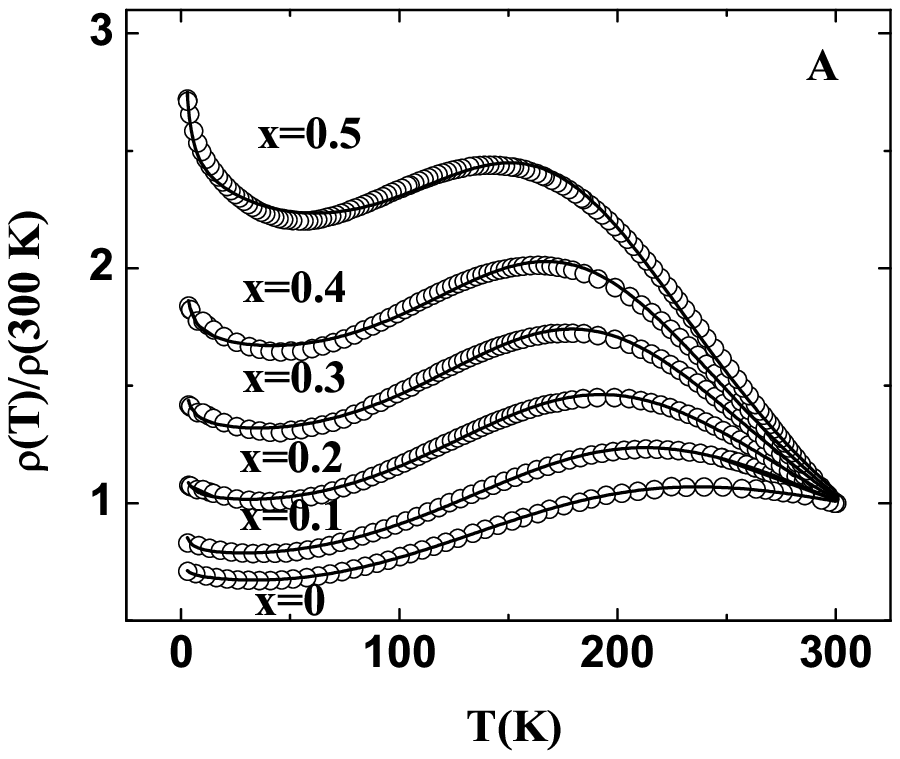}}
\resizebox {4.2cm}{4.2cm}
{\includegraphics{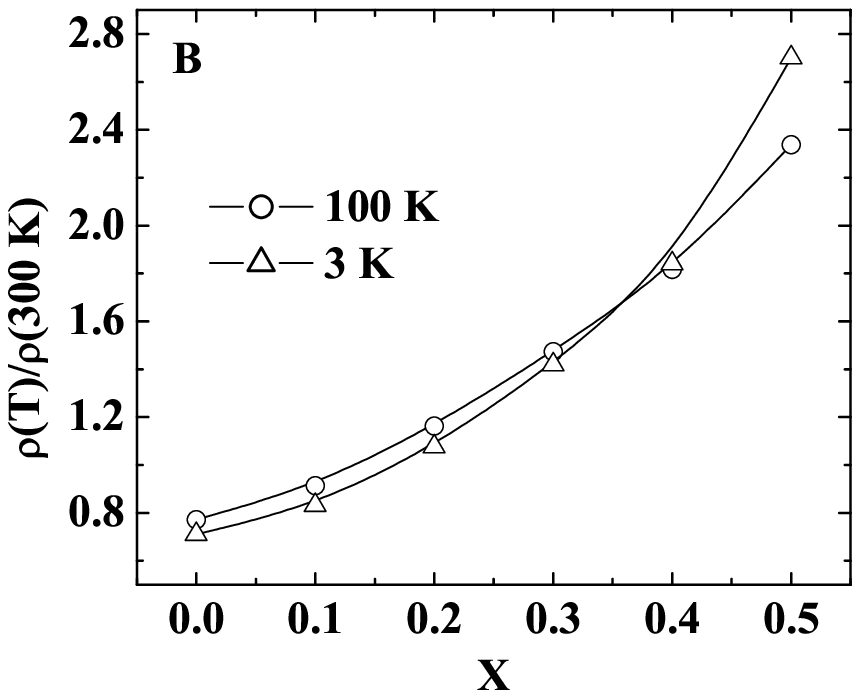}}
%\caption{
%On the left: Resistivity curves for (LSMO)$_{1-x}$(ALO)$_{x}$
%in absence of magnetic
%field along with the corresponding theoritical fits. The open symbols are
%the experimental data and the continuous lines are the theoritical fits.
%On the right: The resistivity as a function of x at $3$ K and $100$ K
%}
%\label{fig:res}
\end{figure}
\centering{Fig:3}
$~~~~~~~~~~~~~$
\vskip 4.5in
\newpage
\begin{figure}
\resizebox {6.5cm}{5.5cm}
{\includegraphics{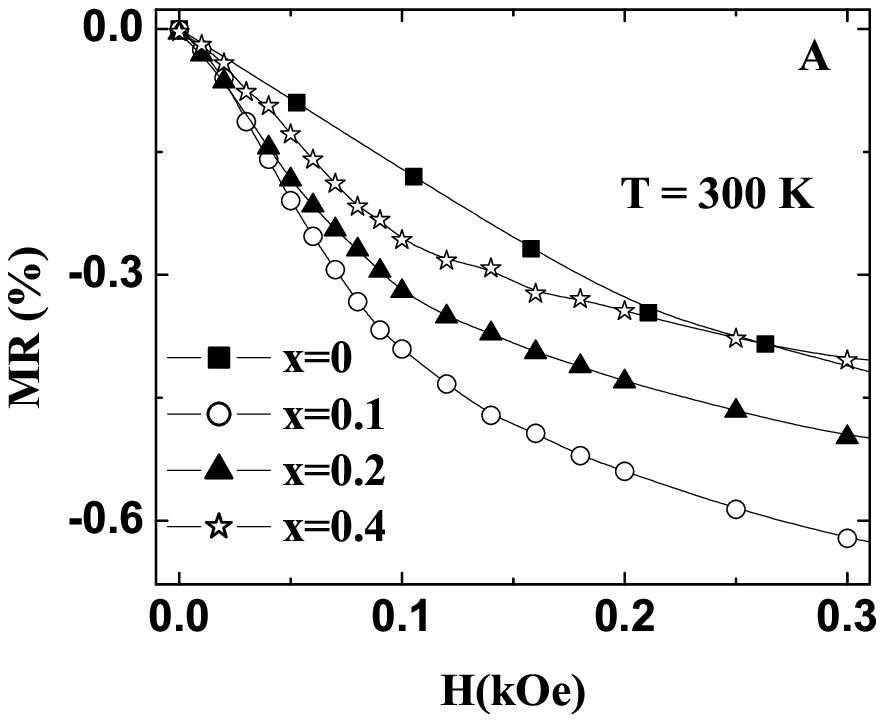}}
\resizebox {6.5cm}{5.5cm}
{\includegraphics{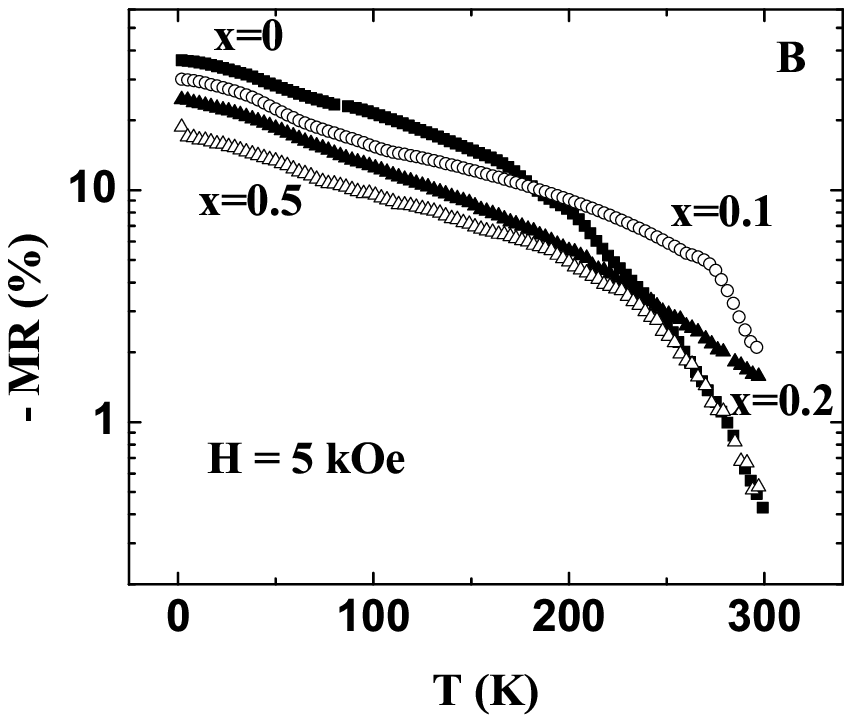}}
%\caption{
%Above: The magnetoresistance as a function of magnetic field for
%different values of $x=0, 0.1, 0.2, 0.4$ at room temperature.
%Enhancement of low field room temperature
%magnetoresistance is observed with introduction of ALO.
%Below: The temperature dependence of magnetoresistance for
%$x=0, 0.1, 0.2, 0.5$ at $H = 5$ kOe. Enhancement of magnetoresistance is
%observed with introduction of ALO near room temperature but at low
%temperature, the enhancement is not observed. The curves corresponding
%to other samples are not shown for clarity.
%}
%\label{fig:mr}
\end{figure}
\centering{Fig:4}

\end{document}